\documentclass[12pt,preprint]{aastex}
\usepackage{url}
\usepackage{rotating}

\shorttitle{Excitation of acoustic waves by vortices in the quiet Sun}

\title{Excitation of acoustic waves by vortices in the quiet Sun}

\author{I.~N. Kitiashvili$^{1}$, A.~G. Kosovichev$^{1}$, N.~N. Mansour$^2$, A.~A. Wray$^2$}
\affil{$^1$ Hansen Experimental Physics Laboratory, Stanford University, Stanford, CA 94305, USA \\
$^2$NASA Ames Research Center, Moffett Field, Mountain View, CA 94040, USA}
\altaffiltext{1}{e-mail: irinasun@stanford.edu}

\begin{document}
\begin{abstract}
Five-minutes oscillations is one of the basic properties of solar convection. Observations show mixture of a large number of acoustic wave fronts propagating from their sources. We investigate the process of acoustic waves excitation from the point of view of individual events, by using realistic 3D radiative hydrodynamic simulation of the quiet Sun. The results show that the excitation events are related to dynamics vortex tubes (or swirls) in the intergranular lanes. These whirlpool-like flows are characterized by very strong horizontal velocities (7 -- 11~km/s) and downflows ($\approx 7$~km/s), and are accompanied by strong decreases of the temperature, density and pressure at the surface and in a $\sim0.5-1$~Mm deep layer below the surface. High-speed whirlpool flows can attract and capture other vortices. According to our simulation results, the processes of the vortex interaction, such as vortex annihilation, can cause the excitation of acoustic waves.

\end{abstract}

\section{Introduction}

Oscillatory behavior is one of the basic properties of the solar surface. Stochastic wave excitation in the turbulent convective medium and frequentative interference of the wavefronts make the problem of solar oscillations very complicated. In observational data, individual wave excitation events are overlaped with other waves.  Nevertheless, individual impulsive events of acoustic waves excitation have been detected by the Vacuum Tower Telescope (Sacramento Peak Observatory). According to the data analysis and the proposed numerical model, the source of solar acoustic waves with a typical period of 5 min is located in the turbulent convection layer $\sim200$~km below the solar surface \citep{Goode1992}. Such impulsive events usually occur in the intergranular lanes of granulation, and are associated with locally high cooling \citep{Rimmele1995}.

At present time there is no clear explanation of the mechanism of acoustic wave sources on the Sun. In the turbulence theory, the wave excitation process was initially studied by \cite{Lighthill1952} for the waves generation in air flow. According to theory, the process of transformation of the flow kinetic energy to the acoustic energy can be caused by short-time deformation of fluid elements by the influence of a shear flow \citep{Lighthill1954}. The importance of turbulence for the wave excitation on the Sun was confirmed by numerical simulations. In particular, the numerical results of \cite{stein2001} supported the possibility of the near surface location of the acoustic sources. The simulations also supported the suggestion that the sources are located  in the intergranular lanes or in their vicinity, where the wave excitation is related to occasional, strong pressure fluctuations associated with the initiation of dowdrafts, and, sometimes with the intergranular lane formation. However, numerical simulations cannot resolve turbulent scales in the solar conditions, therefore accurate sub-grid scale modeling of the turbulence is important. In particular, \cite{jacoutot08a} studied several turbulence models, and found that the best agreement of the synthetic oscillation power spectrum with observational data is given by the dynamic formulation of the Smagorinsky model \citep{Germano1991,Moin1991}.

The problem of resolving small-scale turbulent effects exists also in the observations. Recent observations were able to resolve relatively large, $\sim 3$~Mm in diameter, vortex flows in the photosphere \citep{brandt1988}. Previously, an evidence for vortex motions was found from an example of two bright points rotating around each other \citep{wang1995}. Other observations of vortices showed connection of the vortex motions to strong downflows \citep{potzi2005}. Also, the observations were able to detect a number of vortical flows in a quiet Sun region, near the solar disk center, by tracing  magnetic bright points \citep{bonet08}, which followed a logarithmic spiral trajectory around intergranular points and were engulfed by a downdraft. These whirlpools have the size $\leq0.5$~Mm, and their lifetime varies from 5~min to $\geq 20$~min \citep{bonet08,bonet10}. The search and identification problem of vortices in solar data is mainly the problem of spatial resolution of the observations. The distribution of vortices shows a strong preference to concentration in regions of convective downflows, particulary, at the mesogranular scale \citep{potzi2005}.

In this paper, we present new results of 3D radiative hydrodynamics numerical simulations of a quiet Sun region, identify the process of excitation of acoustic waves, and study their propagation in the convective medium. Our results reveal an interesting connection between a process of interaction of vortex tubes, distributed in the near-surface layer of the Sun, and the generation of acoustic waves.

\section{Realistic simulations of the quiet Sun turbulent convection: numerical setup}

We used the 3D radiative MHD simulation code ("SolarBox", developed by A. Wray). It is based on a Large Eddy Simulations (LES) formulation and includes various subgrid-scale turbulence models \citep{jacoutot08a}. This code takes into account several physical phenomena: compressible fluid flow in a highly stratified medium, three-dimensional multi-group radiative energy transfer between the fluid elements, a real-gas equation of state, ionization and excitation of all abundant species. The code has been carefully tested, and used for studying how various turbulence models affect the excitation of solar acoustic oscillations by turbulent convection in the upper convection zone \citep{jacoutot08a,jacoutot08b}. The code was verified by comparison with the code of \cite{stein2001}, and has been used to study solar oscillations \citep{jacoutot08a}, magnetoconvection of sunspots \citep{kiti2009}, and processes of spontaneous formation of magnetic structures \citep{kiti2010}.

The simulation results are obtained for computational domains of $6.4\times6.4\times5.5$~Mm$^3$ and $12\times12\times5.5$~Mm$^3$, and various grid space: $50^2\times43$~km$^3$, $25^2\times21.7$~km$^3$ and $12.5^2\times11$~km$^3$. The domains include a 5~Mm-deep layer of the upper convective zone and a 0.5~Mm-high layer of the low atmosphere. The lateral boundary conditions are periodic. The top and bottom boundaries are closed  to mass, momentum and energy transfer (apart from radiative losses).

\section{Vortex structures in the simulated quiet Sun}
Vortical motions in the intergranular lanes ("inverted tornadoes") initially were predicted by numerical simulations \citep{nordlund1985,stein2000}, which also showed correlation of the inverted tornadoes.
Figure~\ref{2D} shows snapshots of the surface layer for the temperature, density, vertical velocity and the magnitude of horizontal velocities for the same moment of time. The white squares indicate several vortex structures (whirlpools). The figure also demonstrates difficulties of detection small vortices in the temperature and vertical velocity distributions. The detailed structure of a single vortex is shown in the bottom left corner of each panel vortices in Figure~\ref{2D}. The vortices have a complicate structure that can be described by the following properties: i) a low-temperature core; ii) dramatically decreased density and gas pressure; iii) strong (up to 7~km/s) down flows; and iv) high-speed, often supersonic, horizontal velocities that reach up to 11~km/s.

\section{Annihilation of vortices as a source acoustic waves}

\begin{figure}
\includegraphics[width=1\linewidth]{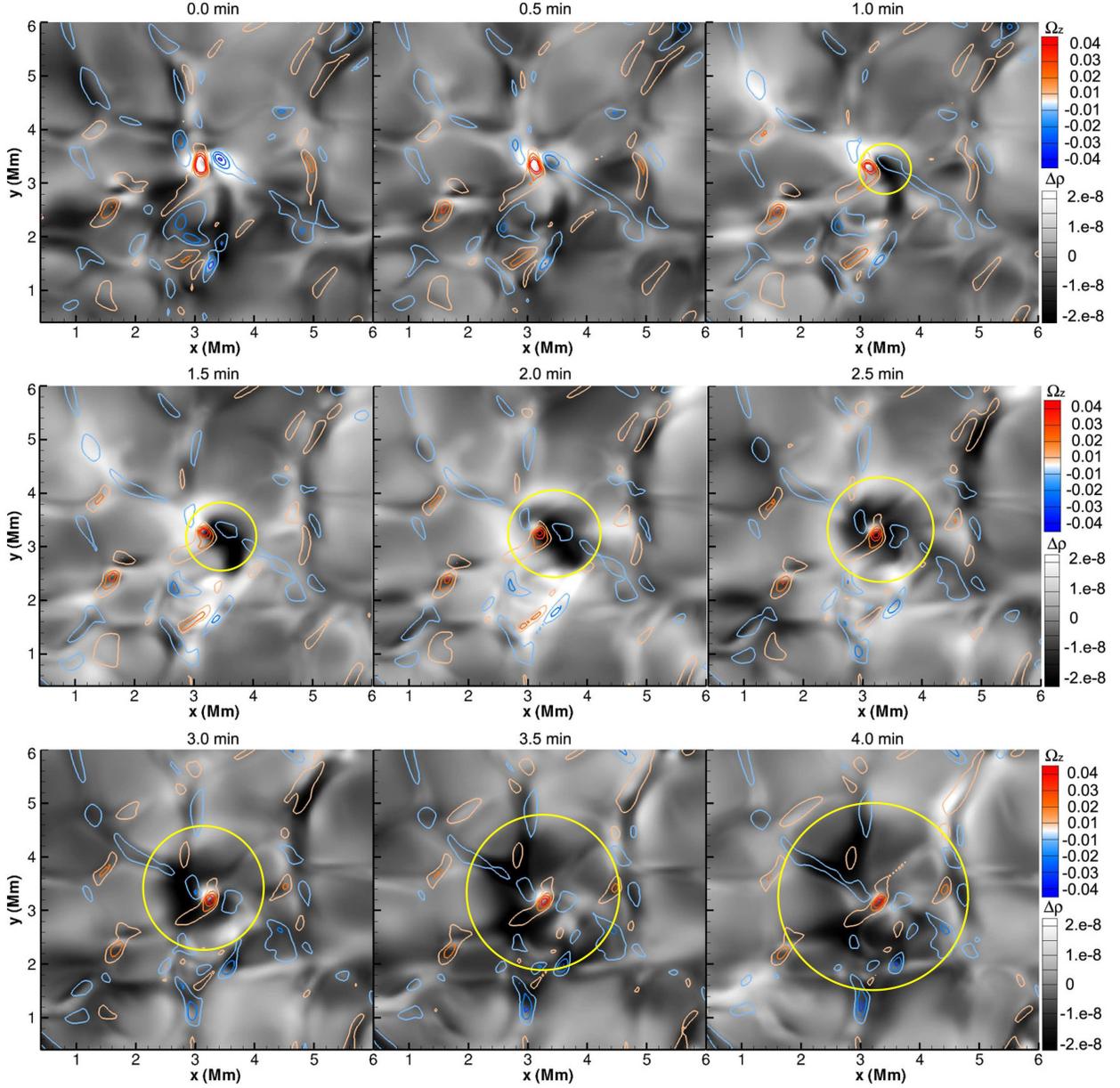}
\caption{Temporal evolution of density fluctuations (calculated as $\rho(t_{i+1})-\rho(t_{i})$) at the solar surface with cadence 30~sec shows acoustic wave excitation and radial propagation from a vortex source, representing interaction of two vortices with opposite-sign vertical vorticity, $\Omega_z$. Overplotted yellow circles indicate approximate position of the wave front. Red and blue lines correspond to the magnitude of the positive (clockwise) and negative (counterclockwise) vertical vorticity. \label{2Dwaves}}
\end{figure}

The dynamics of vortices in the intergranular lanes resembles a "tornado"-like behavior. An accurate tracking of individual vortices reveals concentric waves, excited approximately at the center of the strong vortex cores (see movie\footnote{ \url{http://sun.stanford.edu/~irina/papers/wave/AcousticWave.avi}}). The waves are best seen in density perturbations and can be a result of interaction two or more vortices. The wave fronts have a form of a circle or a sector, depending on the source structure, the interference of waves from other sources and on the dynamic of surrounding flows. The wave propagation can be observed also in the movies of the vertical velocity, temperature and intensity variations at the surface. Figure~\ref{2Dwaves} shows the process of the acoustic wave excitation in a sequence of the density difference, $\rho(t_{i+1})-\rho(t_{i})$, with the cadence of 30~sec, in a horizontal slice corresponding to the surface. Yellow circles indicate the approximate positions of the wave front. The wave originated from a place of collision of two vortices, the vortical vorticity of which had opposite signs (indicated by red and blue contours).

This process of acoustic wave excitation is common in our simulations. However, the identification of individual events is often difficult because the wave amplitude is of the same order as the amplitude of noise coming from other acoustic waves and turbulent convection. Nevertheless, we identified and examined many excitation events with circular-shaped wave fronts, similar to shown in Figure~\ref{2Dwaves}.

\begin{figure}
\begin{center}
\includegraphics[scale=0.9]{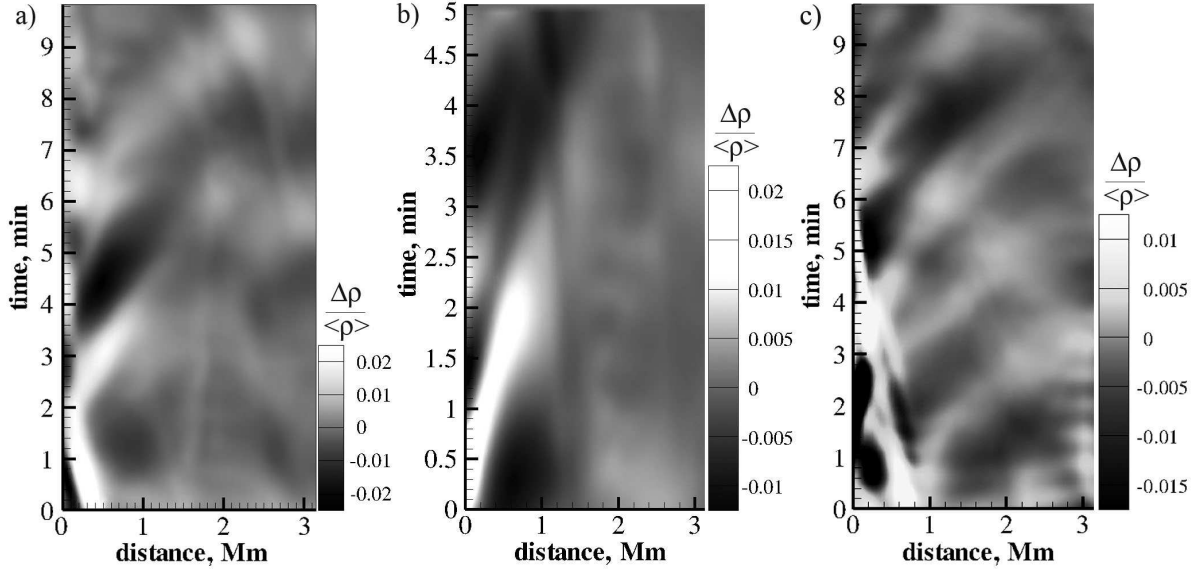}
\end{center}
\caption{Time-distance diagrams of the normalized density fluctuations show inclined ridges, corresponding to acoustic waves.Panel a) correspond to the event shown in Fig.~\ref{2Dwaves}. The slope of the ridges corresponds to a mean speed of 7 -- 14~km/s. \label{time-dist}}
\end{figure}

The propagating wave fronts can be shown also on time-distance diagrams, obtained by plotting the density perturbations averaged over a range of angles for different distances from the vortex locations. Figure~\ref{time-dist} shows three examples of acoustic waves propagation in such time-distance diagrams. The diagrams show the normalized density variations. The light inclined ridges correspond to the acoustic waves. The wave speed is on average 7 -- 14~km/s, but may vary probably due the background convective flows. In some cases, we see, sequences of acoustic waves (Figure~\ref{time-dist}a and c), when several wave fronts are produced with $\simeq 2-2.5$~min intervals. Our analysis shows that occurence of the excitation events mostly depends on the distribution of vortices and their dynamics.

\begin{figure}
\begin{center}
\includegraphics[scale=0.8]{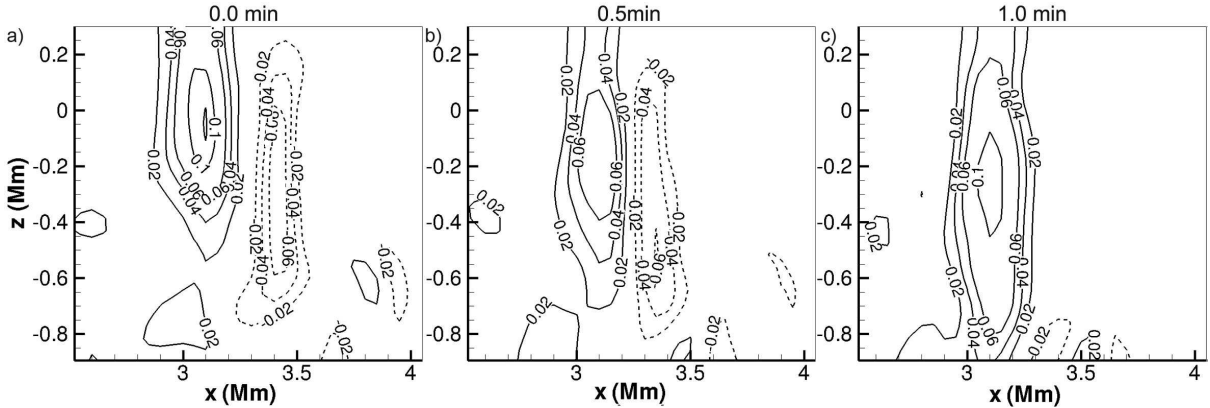}
\end{center}
\caption{Subsurface interaction of the vortices shown in  at different stages: a) initial structure of the vortices, b) closing-up stage, and c) after partial annihilation. Solid and dashed isolines show the  magnitude of the positive and negative vertical vorticity ($s^{-1}$).  \label{vortices-xz}}
\end{figure}

\begin{figure}
\begin{center}
\includegraphics[scale=1.]{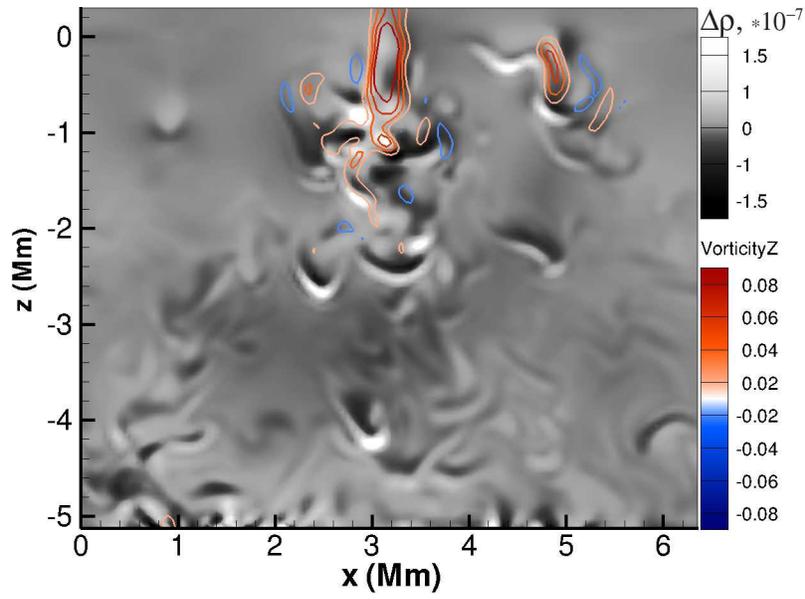}
\end{center}
\caption{Acoustic waves propagation through the subsurface layers. The grayscale background shows the density difference. Solid and dashed isolines are positive and negative vertical vorticity contours. \label{wave-xz}}
\end{figure}

The vertical vorticity distribution reveals a correlation between the excitation events of the acoustic waves and the interaction of vortices with the opposite-sign vorticity, similar to shown in Figure\ref{2Dwaves}. In this particular example, two local concentrations of the vertical vorticity with the opposite signs move close to each other and partially annihilate under the surface, resulting in a partial cancelation (magnitude reduction) of the negative vorticity. This process produces a strong negative density perturbation at the surface (Fig.~\ref{2Dwaves}) and in the subsurface layers. Figure~\ref{vortices-xz} shows different stages of the process of the subsurface interaction of the vortices. In this process, two vortices that rotate in the opposite directions (panel a) move closer to each other (panel b) and partially annihilate (panel c). The annihilation of vortices is often partial, without complete vanishing of the smaller vortex. In general, the vortex interaction is quite complex: turbulent flows deform the shape of the swirls and also stretch smaller vortexes around bigger "main" vortices. The vortex annihilation process can lead to strong local density perturbations. Figure~\ref{wave-xz} illustrates the propagation of an acoustic wave through the subsurface layers.

The depth of the wave excitation events (or the annihilation process) is within the top 0.5~Mm layer. But, more detailed investigation is needed, because the complicated structure and dynamics of the vortex tubes result in very complicated wave fronts.

\section{Discussion and Conclusion}
\begin{figure}
\begin{center}
\includegraphics[scale=1]{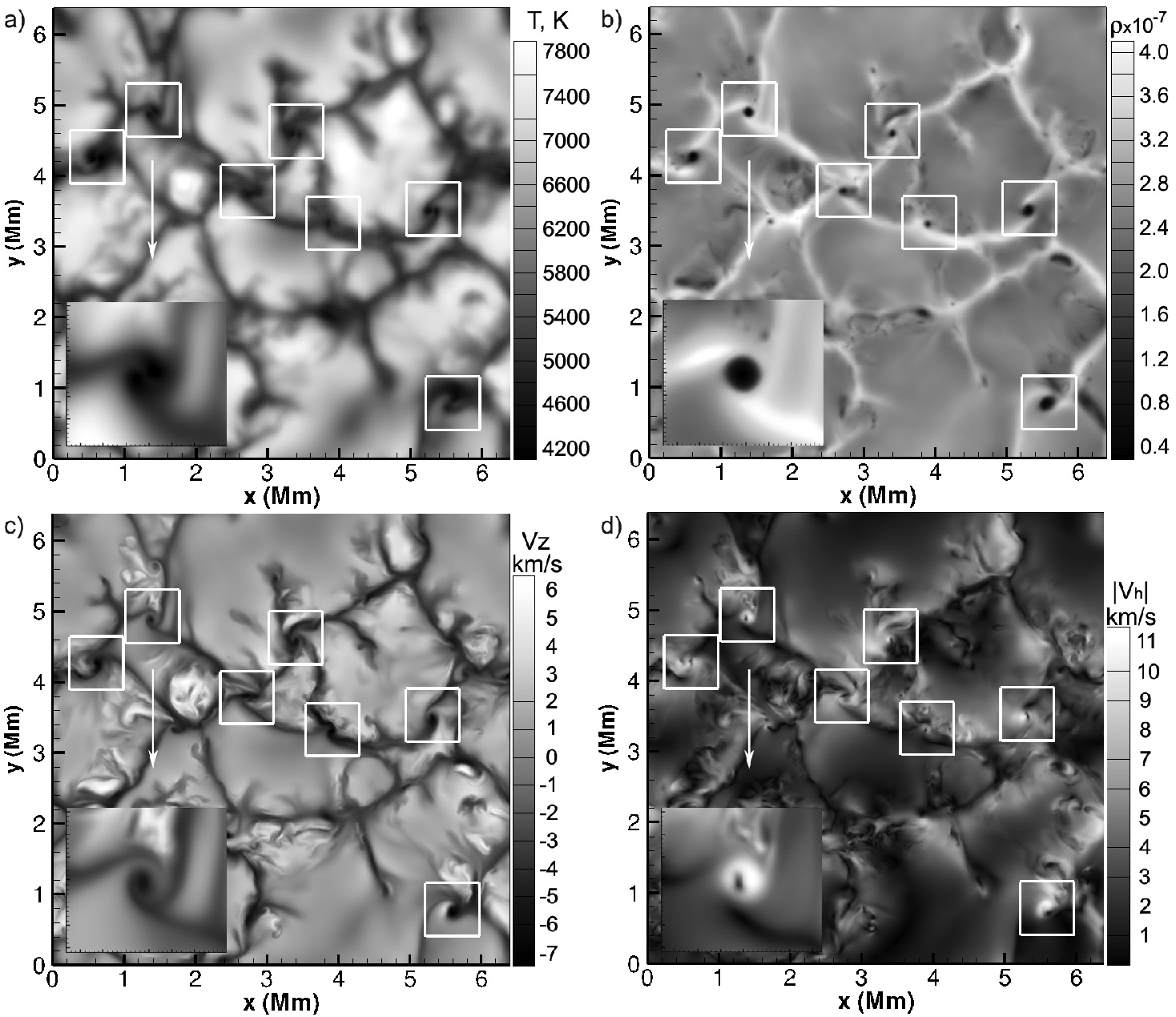}
\end{center}
\caption{Snapshots of the simulated solar surface for a) temperature, b) density, c) vertical velocity and d) magnitude of the horizontal velocities. White squares indicate the largest whirlpools. The whirlpools are located in the intergranular lanes. One of the whirlpool is magnified and shown in the left bottom corner of the each panel. The horizontal grid resolution in this simulation run is 12.5 km. \label{2D}}
\end{figure}

There is no doubt that the solar oscillations are a result of the highly turbulent convective flows in the stratified subsurface layers. The results of our simulations reveal strong interacting vortices in the top layers of the convection zone. The vortices are often characterized by supersonic horizontal flows and strong downflows in the vortex cores (Fig.~\ref{2D}).  The simulation results do not show a preference in directions of the vortex rotation. The vortices are formed in the intergranular lines. They are numerous and interact with each other. Such vortical behavior in the subsurface layers is very common.

We have found that the process of interaction of the vortices with opposite-sign vorticity can lead to their partial annihilation, and that this can produce strong density perturbations and excitation of acoustic waves. The dynamics of the vortices is complicated and mostly depends on the structure of surrounding convective motions. Our future plan is to investigate physical properties of the acoustic sources quantitatively, and compare with helioseismology observations.

\newpage

\end{document}